\documentclass[pra,twocolumn,showpacs,groupedaddress,superscriptaddress]{revtex4-1}
\usepackage{bm,graphicx,amsmath}
\usepackage{placeins}
\usepackage{amssymb}
\usepackage{amsmath}
\usepackage{pstricks}
\usepackage{epsfig}
\newcommand{\fr}{{\rm f}}
\newcommand{\br}{{\rm b}}
\newcommand{\refeqs}[1]{Eqs.~(\ref{#1})}

\begin{document}
\title{Quantum state storage and processing for polarization qubits \\ 
in an inhomogeneously broadened $\Lambda$-type three-level medium}

\date{\today}

\author{D. Viscor}
\affiliation{Departament de F\'{\i}sica, Universitat Aut\`{o}noma de Barcelona, E-08193 Bellaterra, Spain } 
\author{A. Ferraro}
\affiliation{Departament de F\'{\i}sica, Universitat Aut\`{o}noma de Barcelona, E-08193 Bellaterra, Spain } 
\author{Yu. Loiko}
\affiliation{Departament de F\'{\i}sica, Universitat Aut\`{o}noma de Barcelona, E-08193 Bellaterra, Spain } 
\affiliation{Institute of Physics, National Academy of Sciences of Belarus, Nezalezhnasty Ave. 68, 220072 Minsk, Belarus}
\author{J. Mompart}
\affiliation{Departament de F\'{\i}sica, Universitat Aut\`{o}noma de Barcelona, E-08193 Bellaterra, Spain } 
\author{V. Ahufinger}
\affiliation{Departament de F\'{\i}sica, Universitat Aut\`{o}noma de Barcelona, E-08193 Bellaterra, Spain } 

\begin{abstract}
We address the propagation of a single photon pulse with two polarization components, i.e., a polarization qubit, in an inhomogeneously broadened ``phaseonium'' $\Lambda$-type three-level medium.
We combine some of the non-trivial propagation effects characteristic for this kind of coherently prepared systems and the controlled reversible inhomogeneous broadening technique to propose several quantum information processing applications, such as a protocol for polarization qubit filtering and sieving as well as a tunable polarization beam splitter. Moreover, we show that, by imposing a spatial variation of the atomic coherence phase, an efficient quantum memory for the incident polarization qubit can be also implemented in $\Lambda$-type three-level systems.
\end{abstract}


\maketitle

\section{Introduction}\label{intro}

The propagation of electromagnetic pulses in multilevel media has been widely investigated in the last decades. Three level atomic media interacting with two optical fields in a $\Lambda$-type configuration have been among the most considered systems, leading to the discovery of a large variety of phenomena, such as coherent population trapping, electromagnetically induced transparency (EIT), and slow light \cite{SZ}. More recently, the development of quantum technologies for quantum information applications has triggered a renewed attention on the subject. In particular, the propagation of weak (quantum) pulses in a $\Lambda$ medium in the presence of strong (classical) driving fields has been considered in detail, giving rise to proposals and implementations for quantum state storage and processing \cite{nature_review}. 

The non-absorbing propagation of a pair of pulses in coherently prepared $\Lambda$-type media has been deeply investigated in relation to pulse matching \cite{Harris'93,Harris'94,Fleischhauer'95,Kozlov'00}, the dark area theorem \cite{Eberly'02}, simultons \cite{Konopnicki'81} and adiabatons \cite{Grobe'94,Eberly'94}. In all these investigations, the initial quantum state of the atomic population plays an essential role. The term ``phaseonium'' was introduced by Scully \cite{Scully'92} to describe a coherent pure-state superposition between the ground levels of a $\Lambda$-medium and methods to prepare a phaseonium state have been put forward for ladder \cite{Sangouard'06} and $\Lambda$ \cite{Kozlov'09} type systems. An interesting effect that takes place in these systems is the loss-free propagation of light pulses in otherwise optically thick media. In particular, it has been shown that matched pulses, i.e., pulse pairs with identical envelopes, can propagate without distortion if the atoms are prepared in a suitable phaseonium state \cite{Harris'94}.

Particularly relevant for the purposes of the present study is the propagation of two weak optical pulses through an inhomogeneously broadened phaseonium medium. For this system, it was shown in Ref.~\cite{Fleischhauer'95} that, under the two-photon resonance condition, a certain superposition of the fields ---the {\it antisymmetric} normal mode \cite{Harris'94,Eberly'94}--- does not couple to the coherent atomic state, whereas the orthogonal superposition ---{\it symmetric} normal mode--- does interact with the atoms and is completely absorbed. Both the antisymmetric and symmetric modes are determined by the phaseonium state, and can therefore be tuned according to the phase of the atomic coherences.  

In this article, we consider the combination of the above mentioned propagation effects occurring in a $\Lambda$-type three level medium interacting with two weak pulses together with the controlled reversible inhomogeneous broadening (CRIB) \cite{crib} technique, first introduced by Moiseev and Kr\"{o}ll \cite{MK} for quantum state storage. In the original CRIB proposal \cite{crib}, a Doppler broadened atomic vapor was considered for storage and retrieval of a single photon, exploiting the time-reversal symmetry of the optical Bloch equations. Later on, this seminal idea was extended to solid state systems with a twofold advantage: long-living metastable states allow for longer storage times and the inhomogeneous broadening can be  artificially controlled \cite{SanRC,sanrep,polmem_solid,peqm_review}. 
CRIB techniques have been originally developed for systems composed of two-level atoms with the aim of storing quantum information encoded in time domain ---so called time-bin qubits. However, many quantum information processes and sources of photon states are based on the polarization degree of freedom, showing a high degree of interferometric stability and experimental compactness. Most proposals addressing the storage and retrieval of polarization qubits have considered the possibility of spatially splitting the original qubit and storing each polarization component separately \cite{polmem_ensembles}. Only recently, it has been discussed the implementation of polarization quantum memories that do not rely on the spatial splitting of each polarization component of single photons~\cite{Specht'11,Carreño'10,Viscoretal}.

In this work, we will show that the CRIB technique offers also the possibility to store and retrieve quantum states of polarization qubits in a coherently prepared $\Lambda$-type system with ideally perfect efficiency and fidelity. We notice here that the application of the CRIB technique to this kind of systems is not straightforward since, as already mentioned, part of both polarization components of the single photon will propagate as matched pulses and, therefore, it will not be absorbed. Hence, the photon quantum information can not be, in general, perfectly stored in a $\Lambda$-type medium. 
Taking profit of this fact, we will show that tunable polarization filters and sieves can be devised, as well as generic tunable polarization splitters. 
Moreover, we will show that, by forcing the field components to propagate without the possiblity of adjusting their amplitudes as matched solitons, an efficient quantum memory can also be implemented in three-level $\Lambda$ media.

The paper is organized as follows. 
In Sec.~\ref{sec:Equations} we describe the physical system and write down the optical Bloch equations that govern its evolution. We obtain the solutions for the incident field components in Section~\ref{sec:TuPoQuS} and show that, by applying the CRIB technique, the system can be used to implement a tunable polarization qubit splitter.
Next, in Section~\ref{sec:QM}, we consider the case for which the phase of the coherent preparation of the medium is position dependent. For this particular case we demonstrate that, for a large enough optical depth, both field components can be efficiently absorbed and retrieved preserving the amplitude and relative phase between them.
The validity of the assumptions made for the analytical calculations are verified in Section~\ref{sec:Numerics} by numerically integrating the full optical Bloch equations.
Finally, we summarize the results of this work and present the main conclusions in Section~\ref{sec:Conclusions}.

\section{The model}
\label{sec:Equations}

\begin{figure}
\includegraphics[width=\columnwidth]{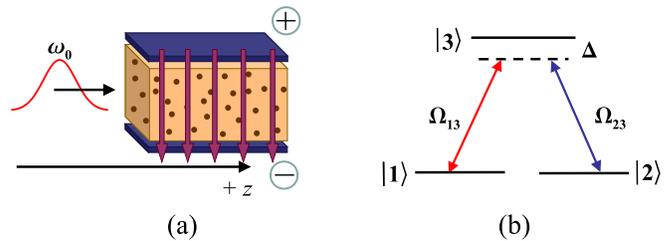}
\caption{(Color online) (a) Physical system under investigation:
a single pulse with central frequency $\omega_0$ and two
circular polarization components enters
a medium with transverse inhomogeneous broadening.
(b) Scheme of the $\Lambda$-type interaction between the single pulse polarization components and the three-level atoms of the phaseonium medium. 
$\Omega_{13}$ and $\Omega_{23}$ are the Rabi frequencies of the corresponding interaction and $\Delta $ is the two-photon detuning.}
\label{f:figs1ab}
\end{figure}

We consider the physical system sketched in Fig.~\ref{f:figs1ab}. A single pulse, with central frequency $\omega_{0}$ and two orthogonal polarization components
(for definiteness we assume left and right circular polarizations)
propagates in the $z$ direction.
The pulse interacts with a medium in 
which an artificial 
%% transverse
inhomogeneous broadening,
much wider than the spectral width of the pulse, has been created. 
This can be achieved, for instance, via Stark shift of the atom levels by an externally applied electric field.
Left and right circular components of the field interact with the $\left|1\right\rangle-\left|3\right\rangle$ and $\left|2\right\rangle-\left|3\right\rangle$ transitions, respectively, defining a $\Lambda$-type interaction.
The atoms of the medium have been initially prepared in a coherent superposition of the two ground levels $\left|1\right\rangle$ and $\left|2\right\rangle$.
We assume that the two-photon resonance condition is fulfilled for all atoms
within the inhomogeneously broadened profile.

\FloatBarrier
% % % % % % % % % % % % % % % % % % % % % % % % % % % % % % % %
% % % % % % % % % % % % % % % % % % % % % % % % % % % % %
% Equations of the system

The evolution of a single atom of the medium can be described, in the rotating wave and electric dipole approximations, by the following equations:
\begin{subequations} \label{sigmaijeqs}
\begin{eqnarray}
\frac{\partial}{\partial t}\sigma_{11}(z,t)&=& i\sigma_{13}(z,t)\Omega_{13}^{\ast}(z,t)+ {\rm c.c.}, \\
\frac{\partial}{\partial t}\sigma_{22}(z,t)&=& i\sigma_{23}(z,t)\Omega_{23}^{\ast}(z,t)+ {\rm c.c.}, \\
\frac{\partial}{\partial t}\sigma_{12}(z,t)&=& i\sigma_{13}(z,t,\omega_{31})\Omega_{23}^{\ast}(z,t)-i\sigma_{32}(z,t)\Omega_{13}(z,t), 
\notag \\ \label{sigma12eqs} \\
\frac{\partial}{\partial t}\sigma_{13}(z,t)&=& i\sigma_{12}(z,t)\Omega_{23}(z,t)-i\omega_{31}\sigma_{13}(z,t) \notag \\
&&-i\left[\sigma_{33}(z,t)-\sigma_{11}(z,t)\right]\Omega_{13}(z,t) \label{sigma13eqs}, \\
\frac{\partial}{\partial t}\sigma_{23}(z,t)&=& i\sigma_{21}(z,t)\Omega_{13}(z,t)-i\omega_{32}\sigma_{23}(z,t) \notag \\
&&-i\left[\sigma_{33}(z,t)-\sigma_{22}(z,t)\right]\Omega_{23}(z,t) \label{sigma23eqs},
\end{eqnarray}
\end{subequations}
where $\sigma_{ii}$ is the populations of level $\left|i\right\rangle$, $\sigma_{ij}$ is the atomic coherence between levels $\left|i\right\rangle$ and $\left|j\right\rangle$, $\Omega_{ij}=\left(\vec{d}_{ij}\vec{E}_{ij}\right)/\hbar$ is the Rabi frequency
with $\vec{E}_{ij}$ being the slowly varying electric field amplitude of the light component coupled with transition $\left|i\right\rangle \leftrightarrow \left|j\right\rangle$, $\vec{d}_{ij}$ is the dipole moment of the corresponding transition, $\hbar$ is the reduced Planck constant, and $\omega_{ji}=\omega_{j}-\omega_{i}$ is the transition frequency between levels $\left|i\right\rangle$ and $\left|j\right\rangle$. Note that we consider a 
closed atomic system satisfying $\sigma_{11}+\sigma_{22}+\sigma_{33}=1$. Moreover, for simplicity, we have not included any incoherent decay term into Eqs.~(\ref{sigmaijeqs}), since we assume that the lifetimes of the excited level and of the ground state coherence are much larger than the whole duration of the interaction process.

In three-level media it is common to define a new basis of the system in terms of the bright $\left|B\right\rangle$ and dark $\left|D\right\rangle$ states \cite{Fleischhauer'96,ArimondoBook'96} which, under the two-photon resonance condition, are given by
% Bright-Dark states:
\begin{subequations} \label{BDStates}
\begin{eqnarray} 
\left|B\right\rangle &\equiv& \frac{1}{\Omega_{T}} \left(\Omega_{13}\left|1\right\rangle + \Omega_{23}\left|2\right\rangle\right){\rm \ and}\\
\left|D\right\rangle &\equiv& \frac{1}{\Omega_{T}} \left(\Omega^{*}_{23}\left|1\right\rangle - \Omega^{*}_{13}\left|2\right\rangle\right),
\end{eqnarray} 
\end{subequations}
where $\Omega_{T}=\sqrt{\left|\Omega_{13}\right|^{2}+\left|\Omega_{23}\right|^{2}}$ is the total Rabi frequency.
For continuous wave light fields, it has been shown that the dynamics is restricted to the transition $\left|B\right\rangle$ $\leftrightarrow$ $\left|3\right\rangle$. In this case, once the system enters the dark-state due to e.g., spontaneous emission from the excited level, it becomes trapped.
However, for light pulses the dynamics is different. Indeed, Kozlov and Eberly \cite{Kozlov'00,Eberly'02} showed that the dark-state completely controls the system evolution at large propagation distances when the medium is in a phaseonium preparation \cite{Scully'92}.

In order to solve analytically the evolution equations of the system
we follow the treatment given in Ref.~\cite{sanrep} and \cite{Viscoretal},
where the transition operators and the field amplitudes associated to each dipole transition are split
into the forward and backward modes (denoted by the superscripts $\fr$ and $\br$, respectively):
\begin{subequations}
\begin{eqnarray} 
\sigma_{\mu\rho}(z,t)&=&\sigma_{\mu\rho}^{\fr}(z,t)e^{-i(\omega_{0}t-k_{0}z)} \notag \\ 
&+&\sigma_{\mu\rho}^{\br}(z,t)e^{-i(\omega_{0}t+k_{0}z)} \label{sigmaf-b}, \\
\Omega_{\mu\rho}(z,t)&=&\Omega_{\mu\rho}^{\fr}(z,t)e^{-i(\omega_{0}t-k_{0}z)} \notag \\
&+&\Omega_{\mu\rho}^{\br}(z,t)e^{-i(\omega_{0}t+k_{0}z)},  \label{fieldf-b}
\end{eqnarray}
\end{subequations}
where $\omega_0$ is the central frequency of the pulse and $k_0$ the corresponding wavenumber.
From now on, we set $\rho=3$, $\mu,\nu=1,2$ and $\mu\neq\nu$. We assume that the initial population is distributed only between levels $\left\vert 1\right\rangle$ and $\left\vert 2\right\rangle$,
with the excited level $\left\vert 3\right\rangle$ being unpopulated.
Since the interaction involves weak pulses,
one can neglect \cite{crib} the temporal variations in the level populations 
as well as in the coherence $\sigma_{12}$.
The validity of these approximations
is checked and confirmed in Sec.~\ref{sec:Numerics} via numerical analysis of the full set of
the Maxwell-Bloch Eqs.~(\ref{sigmaijeqs}) and (\ref{fieldeqf}). 

Using the forward/backward mode decomposition ((\ref{sigmaf-b}) and (\ref{fieldf-b})) and 
the weak field approximation (outlined just above)
Eqs.~(\ref{sigmaijeqs}) are simplified as follows: 
\begin{eqnarray} \label{sigma13eq}
\frac{\partial}{\partial t}\sigma_{\mu\rho}^{\br,\fr}(z,t,\Delta)&=& i\Delta\sigma_{\mu\rho}^{\br,\fr}(z,t,\Delta)+i\sigma_{\mu\mu}\Omega^{\br,\fr}_{\mu\rho}(z,t) \notag\\ &&+i\sigma_{\mu\nu}(\phi_{\mu\nu})\Omega^{\br,\fr}_{\nu\rho}(z,t),
\end{eqnarray}
where $\Delta=\omega_{0}-\omega_{31}=\omega_{0}-\omega_{32}$. Notice that we have included an explicit dependence on the phase $\phi_{\mu\nu}=\phi_{\mu}-\phi_{\nu}$ between the two ground levels through the coherence $\sigma_{\mu\nu}$.
This allows to take into account the general case in which the system is initially prepared in an arbitrary coherent superposition of the ground levels. Note that, from Eqs.~(\ref{sigma13eq}) it is clear that for atoms prepared initially in an incoherent mixture of the ground states, $\left|\sigma_{\mu\nu}\right|=0$, the equations for each transition of the lambda system become decoupled. This decoupling also occurs for a polarization qubit memory using $V$-type three level atoms as it was recently reported in Ref.~\cite{Viscoretal}.

The propagation of the forward and backward modes of the light, in a reference frame moving with the pulses, can be described by the following equations:
\begin{eqnarray} \label{fieldeqs}
\frac{\partial}{\partial z}\Omega^{\br,\fr}_{\mu\rho}(z,t) &=&\mp i\eta_{\mu\rho}\int^{\infty}_{-\infty}d\Delta\times \notag \\ 
&\times& \left[G_{\rho\mu}\left(\Delta\right)\sigma_{\mu\rho}^{\br,\fr}(z,t,\Delta)\right], \label{fieldeqf}
\end{eqnarray}
where $+$ ($-$) refers to backward (forward) modes, $G_{\rho\mu}\left(\Delta\right)$ is the inhomogeneous frequency distribution of atoms of the corresponding transition; $\eta_{\mu\rho}=g^{2}Nd^{2}_{\mu\rho}/\hbar c$ with $g^{2}=\omega_{0}/2\varepsilon_{0}V$ being the coupling constant;
$\varepsilon_{0}$ is the vacuum electric permittivity,
$N
%=\int \rho (\Delta_{ji})d\Delta_{ji}
$ is the number of atoms in the quantization volume $V$,
and $c$ is the speed of light in vacuum. It is worth noting that, although the treatment is performed in the semiclassical formalism, the linearity of the equations ensure the validity of this model also at the single photon level. Thus, in what follows, both the atomic coherences and the fields could be interpreted as classical amplitudes as well as quantum operators.

From (\ref{sigma13eq})-(\ref{fieldeqs}), we note that the time reversed ($t\rightarrow-t$) equations for the forward propagating modes are equal to the backward ones under a sign change in the detunings and in the field amplitudes. This symmetry in the optical-Bloch equations is indeed the basis for the CRIB method. Thus, once the forward propagating input light pulse has been completely absorbed, in order to retrieve the pulse as a time reversed copy of itself, we need to reverse the detuning ($\Delta\rightarrow-\Delta$). This reversing operation can be achieved, for instance, by changing the polarity of the device that creates the inhomogeneous broadening. 
At the same time, one has to apply a position dependent phase matching operation to transform the forward components of the atomic excitations into backward components, so that the retrieved pulse propagates in the backward direction. This phase change can be performed by transfering the atomic coherences back and forth to an auxiliary metastable \cite{crib,Minar'10} state which also allows for longer storage times \cite{spinCoh}. 

\section{Tunable Polarization Qubit Splitter}
\label{sec:TuPoQuS}

In the following we will investigate how the three-level system discussed in the previous section can act as a tunable beam splitter for polarization qubits, a device of clear interest in quantum information processing. In particular, it will be shown that the polarization basis that determines the beam splitter action is defined by the phaseonium state. 

\subsection{Quantum filter}
% Equations
Let us first consider the propagation of the incident pulse in the forward direction.
For a medium of $\Lambda$-type atoms prepared initially in a coherent superposition of the ground levels $\left|1\right\rangle$ and $\left|2\right\rangle$, i.e., a phaseonium state \cite{Scully'92}, 
Eqs.~(\ref{sigma13eq})-(\ref{fieldeqs}) can be analytically solved following Ref.~\cite{sanrep}.
Thus, we insert the solution of Eqs.~(\ref{sigma13eq}) into Eqs.~(\ref{fieldeqs}),
and Fourier transform the result:
\begin{eqnarray}  \label{fieldeqijfin}
\frac{\partial}{\partial z}\widetilde{\Omega}^{\rm in}_{\mu\rho}(z,\omega) &=&-\eta_{\mu\rho}\sigma_{\mu\mu}H_{\rho\mu}(\omega)\widetilde{\Omega}^{\rm in}_{\mu\rho}(z,\omega) \notag \\
&& -\eta_{\mu\rho}\sigma_{\mu\nu}(\phi_{\mu\nu})H_{\rho\mu}(\omega)\widetilde{\Omega}^{\rm in}_{\nu\rho}(z,\omega),
\end{eqnarray}
where we have used the initial conditions $\sigma_{\mu\rho}(z,t=-\infty)=0$ and we have defined
\begin{equation}
H_{\rho\mu}(\omega)=\int^{\infty}_{-\infty} G_{\rho\mu}\left(\Delta\right)\int^{\infty}_{0}e^{i\omega\tau}e^{i\Delta\tau}d\tau d\Delta . \label{Hijhom}
\end{equation}
An analytical compact solution of \refeqs{fieldeqijfin} can be given assuming symmetric transitions, i.e., equal electric dipole moments $(|\vec{d}_{13}|=|\vec{d}_{23}|=d)$ and equal inhomogeneous distributions ($G_{31}\left(\Delta\right)=G_{32}\left(\Delta\right)=G\left(\Delta\right)$) yielding $\eta_{13}=\eta_{23}=\eta$ and $H_{32}(\omega)=H_{31}(\omega)=H(\omega)$, which reads:
% Solucions Absorci Lambda Coherent
\begin{eqnarray} \label{SolAbs}
\widetilde{\Omega}^{\rm in}_{\mu\rho}(z,\omega)&=&
\widetilde{\Omega}^{\rm in}_{\mu\rho}(0,\omega) \left(e^{-\alpha(\omega)z}\sigma_{\mu\mu}+\sigma_{\nu\nu}\right) \notag \\ 
&+& \widetilde{\Omega}^{\rm in}_{\nu\rho}(0,\omega)\left|\sigma_{\mu\nu}\right|e^{i\phi_{\mu\nu}}\left(e^{-\alpha(\omega)z}-1\right),
\end{eqnarray}
where $\alpha(\omega)=\eta H(\omega)$ is the absorption coefficient.
These equations describe, for each frequency of the pulse, the propagation of the two polarization components of the incoming field along the medium. 
Each solution depends on both initial polarization components of the weak pulse $\widetilde{\Omega}_{13}^{\rm in}(z=0,\omega)$ and $\widetilde{\Omega}_{23}^{\rm in}(z=0,\omega)$), which implies that, in general, the information carried by each component is mixed as the pulse propagates through the $\Lambda$-medium, due to the two-photon coherence $\sigma_{12}$.

Moreover, it is easy to see from Eq.(\ref{SolAbs}) that, in general, the fields are not completely absorbed, even for large optical depths $\alpha(\omega) z$.
In particular, the two polarization components of the pulse can only be completely absorbed at large optical depths if the condition 
\begin{eqnarray} \label{AbsCond}
\frac{\widetilde{\Omega}^{\rm in}_{13}(0,\omega)}{\widetilde{\Omega}^{\rm in}_{23}(0,\omega)}=\frac{c_{1}}{c_{2}} ,
\end{eqnarray}
with $c_{\mu}c^{*}_{\mu}=\sigma_{\mu\mu}$ is fulfilled.
However, for arbitrary input pulses both components will change their amplitudes and phases
in such a way that the dark-state becomes populated as they propagate \cite{Kozlov'00}.
If the optical depth of the medium is large enough,
eventually the dark-state will be fully populated,
so the field components will propagate as matched solitons, without any further absorption.
Indeed, it can be seen from Eq.~(\ref{SolAbs}) that the field amplitudes will evolve until the relation
\begin{eqnarray} \label{EITCond}
\frac{\widetilde{\Omega}^{\rm in}_{13}(z,\omega)}{\widetilde{\Omega}^{\rm in}_{23}(z,\omega)}=-\frac{c_{2}^{*}}{c_{1}^{*}} 
\end{eqnarray}
is satisfied. Note that relations (\ref{AbsCond}) and (\ref{EITCond}) determine the symmetric and antisymmetric normal modes, respectively, introduced in \cite{Fleischhauer'95}.
In fact, the time reversal symmetry of Eqs.~(\ref{sigma13eq}) and (\ref{fieldeqs}) is not sufficient alone to guarantee full recovery of the original input state.
Clearly, the unabsorbed component of the input pulse cannot be recovered back.
Specifically, the field that exits the medium ends up in a definite polarization state,
which depends only on the phaseonium preparation.
Assuming a large optical depth $\alpha L\rightarrow\infty$, the normalized intensity of each of the polarization components at the output of the medium $z=L$, from Eq.(\ref{SolAbs}), reads:
\begin{eqnarray} \label{SolIntAbs}
\widetilde{I}^{\rm in}_{\mu\rho}(L,\omega)\equiv \frac{\left|\widetilde{\Omega}^{\rm in}_{\mu\rho}(L,\omega)\right|^{2}}{\widetilde{\Omega}^{\rm in}_{T}(L,\omega)^{2}}=\sigma_{\nu\nu},
\end{eqnarray}
where $\widetilde{\Omega}^{\rm in}_{T}(L,\omega)$ is the total Rabi frequency of the output field, as defined in (\ref{BDStates}).
This result perfectly agrees with \cite{Clader'08}, where it was shown 
that, at the output of the phaseonium medium, the amplitude of the field 
coupled with an optical transition depends on the ground-state amplitude of the opposite transition. Moreover, one realizes that in this situation the relative phase between the output field components $\phi_{\mu\rho}(L)-\phi_{\nu\rho}(L)$, using the definition $\widetilde{\Omega}^{\rm in}_{\mu\rho}=\left|\widetilde{\Omega}^{\rm in}_{\mu\rho}\right|e^{i\phi_{\mu\rho}}$, is simply the phase $\phi_{\mu\nu}$ of the $\sigma_{12}$ coherence.
Considering the input pulse to be a single photon in the generic polarization state
%
% Input state
\begin{eqnarray} \label{PsiIn}
\left|\psi_{in}\right\rangle = a_{L}\left|L\right\rangle + a_{R}\left|R\right\rangle ,
\end{eqnarray}
entering a medium prepared in a coherent superposition
%
% Input state
\begin{eqnarray} \label{PsiAtom}
\left|\psi_{at}\right\rangle = c_{1}\left|1\right\rangle + c_{2}\left|2\right\rangle ,
\end{eqnarray}
we will find at the output the state:
% Output state
\begin{eqnarray} \label{PsiOut}
\left|\psi_{out}^\fr\right\rangle = \left(c_{2}-c_{1}\right)^{*}\left(c_{2}\left|L\right\rangle - c_{1}\left|R\right\rangle\right) .
\end{eqnarray}
with probability
\begin{eqnarray} \label{ProbOut}
P_{out}^\fr&\equiv&\left|\left\langle\psi_{in}\right|\left.\psi_{out}^\fr\right\rangle\right|^{2}= \notag \\ &&\left|\left(\sigma_{11}-\sigma_{12}\right)\left|a_{R}\right|^{2}+\left(\sigma_{22}-\sigma_{21}\right)\left|a_{L}\right|^{2}\right|^{2} .
\end{eqnarray} 
Note that since part of the incident light pulse has been absorbed by the medium, the state (\ref{PsiOut}) is not normalized. A convenient picture of the process is that the preparation of the phaseonium medium fixes the basis for which the incident field is filtered: after the dark-state is filled, only the so called antisymmetric normal mode (\ref{PsiOut}) propagates without absorption. In turn the latter depends only on the phaseonium state, which ideally can be tuned at will. In other words, the medium acts as a tunable quantum filter. 

\subsection{Quantum sieve}

As the pulse components propagate through the phaseonium medium, they adjust themselves to fulfill condition (\ref{EITCond}), which allows absorption-free propagation. However, before the dark-state is completely populated, part of the field (the symmetric normal mode, given by (\ref{AbsCond})) is absorbed by the medium. This information stored in the optical coherences can be retrieved back by using the CRIB technique. In this subsection we study the propagation of the retrieved light pulse in the backward direction, which is caused by the sign change of the detunings and the phase matching operation, performed at time $t=0$, once the non-absorbed part of the field has left the medium \cite{sanrep}. In this situation, the equations for the optical coherences and the Rabi frequencies of the corresponding field components are:
\begin{subequations}
\begin{eqnarray}
	\frac{\partial}{\partial t}\sigma_{\mu\rho}^{\br}(z,t,-\Delta)&=& i\sigma_{\mu\nu}(\phi_{\mu\nu})\Omega^{\br}_{\nu\rho}(z,t)+i\sigma_{\mu\mu}\Omega^{\br}_{\mu\rho}(z,t)
 \notag \\
&&-i\Delta\sigma_{\mu\rho}^{\br}(z,t,-\Delta) \label{sigma13eqbout}, \\ 
	\frac{\partial}{\partial z}\Omega_{\mu\rho}^{\br}(z,t)&=& -i\eta_{\mu\rho}\int^{\infty}_{-\infty} d\Delta\times \notag \\ &&\times\left[G_{\rho\mu}\left(-\Delta\right)\sigma_{\mu\rho}^{\br}(z,t,-\Delta)\right], \label{fieldijeqbout}
\end{eqnarray}
\end{subequations}
again with $\rho=3$, $\mu,\nu=1,2$ and $\mu\neq \nu$. The initial and boundary conditions at the time of switch ($t=0$) are:
\begin{subequations} \label{ICbackward}
\begin{eqnarray}
	\sigma_{\mu\rho}^{\br}(z,t=0,-\Delta)&=&
i\int_{-\infty}^{0} e^{-i\Delta s}\left[\sigma_{\mu\mu}\Omega_{\mu\rho}^{\rm in}(z,s)\right. \notag \\
&+&\left. \sigma_{\mu\nu}(\phi_{\mu\nu})\Omega_{\nu\rho}^{\rm in}(z,s)\right]ds , \label{ICsigma13} \\
	\widetilde{\Omega}_{\mu\rho}^{\br}(L,\omega)&=&0 . \label{ICfield}
\end{eqnarray}
\end{subequations}
The first two initial conditions (\ref{ICsigma13}) are obtained from the solution of the atomic coherences (\ref{sigma13eq}) at time $t=0$, whereas the boundary condition (\ref{ICfield}) derives from the assumption that, at the time when the detuning is reversed, the non-absorbed field has left the medium. The above equations can be solved, as for the absorption stage, inserting the solution of (\ref{sigma13eqbout}) into (\ref{fieldijeqbout}), and Fourier transforming the result. By introducing 
\begin{subequations}
\begin{eqnarray}
F_{\rho\mu}(\omega) &=&\int_{-\infty}^{+\infty}G_{\rho\mu}\left(-\Delta\right)
\int_{0}^{\infty}e^{i{\omega}\tau}e^{-i{\Delta}\tau}d\tau d\Delta \label{Fijhom} , \notag \\ \\
J_{\rho\mu}(\omega) &=&\int_{-\infty}^{+\infty}G_{\rho\mu}\left(-\Delta\right)
\int_{-\infty}^{+\infty}e^{i{\omega}\tau}e^{-i{\Delta}\tau}d\tau d\Delta \label{Jijhom} ,  \notag \\
\end{eqnarray}
\end{subequations}
the equations for the backward fields associated with the two transitions are: 
\begin{widetext}
\begin{eqnarray} \label{back_noph}
	\frac{\partial}{\partial z} \widetilde{\Omega}_{\mu\rho}^{\br}(z,\omega)&=&
	 \eta_{\mu\rho}F_{\rho\mu}(\omega)\left[
\sigma_{\mu\nu}(\phi_{\mu\nu})\widetilde{\Omega}_{\nu\rho}^{\br}(z,\omega)+ \sigma_{\mu\mu}\widetilde{\Omega}_{\mu\rho}^{\br}(z,\omega)\right] \notag \\
&+&\eta_{\mu\rho}J_{\rho\mu}(\omega)\left[ \sigma_{\mu\nu}(\phi_{\mu\nu})\widetilde{\Omega}_{\nu\rho}^{\rm in}(z,-\omega)+ \sigma_{\mu\mu}\widetilde{\Omega}_{\mu\rho}^{\rm in}(z,-\omega)\right]. \label{fieldeq13bout}
\end{eqnarray}
As for the case of absorption, an analytic solution for the above equations can be found. In particular, a compact expression can be derived assuming symmetric transitions and inhomogeneous distributions, as done in the derivation of (\ref{SolAbs}):
\begin{eqnarray} \label{SimultSol}
\widetilde{\Omega}^{\rm b}_{\mu\rho}(z,\omega)=\frac{\eta J(\omega)}{\eta \left(F(\omega)+H(-\omega)\right)} \left(e^{-\eta\left(F(\omega)+H(-\omega)\right)(L-z)}-1\right) \left(\widetilde{\Omega}^{\rm in}_{\mu\rho}(0,-\omega) \sigma_{\mu\mu}+\widetilde{\Omega}^{\rm in}_{\nu\rho}(0,-\omega)\left|\sigma_{\mu\nu}\right|e^{i\phi_{\mu\nu}}\right) .
\end{eqnarray}
\end{widetext}
We note that, similar to the solution for the forward propagating modes (\ref{SolAbs}), the backward reemited components of the field are also a combination of the initial amplitudes. From (\ref{SimultSol}), it is easy to obtain that for a large enough optical depth the normalized intensity at the output surface of the medium ($z=0$) reads:
\begin{eqnarray} \label{NormIntLargeL}
\widetilde{I}^{\rm b}_{\mu\rho}(0,\omega)\equiv \frac{\left|\widetilde{\Omega}^{\rm b}_{\mu\rho}(0,\omega)\right|^{2}}{\widetilde{\Omega}^{\rm b}_{T}(0,\omega)^{2}}=\sigma_{\mu\mu},
\end{eqnarray}
where $\widetilde{\Omega}^{\rm b}_{T}(0,\omega)$ is the total Rabi frequency for the backward components at the output. In the last expression, we have considered that the spectral bandwidth of the pulse is smaller than the inhomogeneous broadening, $F(\omega)\simeq H(-\omega)\simeq J(\omega)/2\simeq\pi/2$, and that the absorption coefficient reduces to $\alpha\simeq\eta \pi/2$. In this situation, the relative phase between the field components corresponds to the phase difference between the ground levels, $\phi_{\mu\rho}(0)-\phi_{\nu\rho}(0) = \phi_{\mu\nu}$. For the single-photon case, the backward retrieved components exit the medium in a state
\begin{eqnarray} \label{PsiBack}
\left|\psi_{out}^\br\right\rangle = -\left(c_{1}+c_{2}\right)\left(c^{*}_{1}\left|L\right\rangle + c^{*}_{2}\left|R\right\rangle\right) ,
\end{eqnarray}
which is orthogonal to (\ref{PsiOut}) and corresponds to the so called symmetric normal mode, with probability 
\begin{equation} \label{ProbBack}
P_{out}^\br=1-P_{out}^\fr ,
\end{equation}
where $P_{out}^\fr$ is given in (\ref{ProbOut}). Therefore, the system acts as a tunable quantum sieve that can reemit the sieved state on demand. Note that since part of the field is absorbed by the medium, the sum of the norms of states (\ref{PsiOut}) and (\ref{PsiBack}) must be one. 
%%%%%%%%%%%%%%%%%%%%%%%
\begin{figure} %{\includegraphics[width=0.5\textwidth,height=0.5\textwidth]{back_forw_eff_uphnoise}}
{\includegraphics[width=\columnwidth]{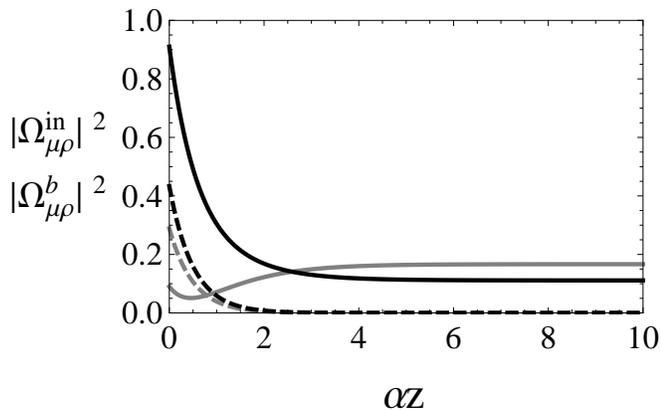}}
\caption{Normalized intensities of the single pulse polarization components (see text for definition) coupled with the $\left|1\right\rangle-\left|3\right\rangle$ (black lines) and $\left|2\right\rangle-\left|3\right\rangle$ (gray lines) optical transitions as a function of the optical distance $\alpha z$. The solid lines correspond to the propagation of the incident field while the dashed lines correspond to the backward retrieved components. The initial prepared state of the phaseonium medium is $\sigma_{11}=0.6$, $\sigma_{22}=0.4$, $\phi_{12}=\pi/3$,
whereas the field components are initially weighed as $I^{\rm in}_{13}(0,t_{c})=0.9$ and $I^{\rm in}_{23}(0,t_{c})=0.1$ with relative phase between them being $\phi_{13}(0)-\phi_{23}(0)=0$.}
\label{f:TuPoQuS}
\end{figure}
%%%%%%%%%%%%%%%%%%%%%%%%

An example of the sieve protocol is shown in Fig.~\ref{f:TuPoQuS},
where the normalized intensity of each polarization component $\left|\Omega_{13}\right|^{2}$ (black lines) and $\left|\Omega_{23}\right|^{2}$ (gray lines) is plotted as a function of the optical distance $\alpha z$. The solid lines correspond to the propagation of the polarization components of the incident field, Eq.~(\ref{SolAbs}), whereas the dashed lines represent the components of the backward retrieved single photon, Eq.~(\ref{SimultSol}).
In this particular case, a medium of optical length $\alpha L=10$ has been chosen with the atomic population prepared initially in a coherent superposition fulfilling $\sigma_{11}=0.6$, $\sigma_{22}=0.4$ and $\phi_{12}=\pi/3$.
The two components of the input light pulse have been assumed to have the relative phase of $\phi_{13}(0)-\phi_{23}(0)=0$ and initial weighs $I^{\rm in}_{13}(0,t_{c})=0.9$ and $I^{\rm in}_{23}(0,t_{c})=0.1$, where $t_{c}$ denotes the temporal position of the pulse peak in the moving frame.
In Fig.~\ref{f:TuPoQuS}, we observe how the incident components change their amplitudes along propagation and that beyond a certain optical length they propagate as matched solitons. The total intensity at the output is given by Eq.~(\ref{ProbOut}), whereas Eqs.~(\ref{SolIntAbs}) determine which fraction of the total transmited intensity is in each mode. Since during the transient regime part of the field has been absorbed, the CRIB technique allows to recover (dashed lines) this information stored in the optical coherences. Clearly, since part of the field has left the medium, the retrieved intensity is smaller than the initial one. In this case, the total retrieved intensity is given by Eq.~(\ref{ProbBack}) while the distribution of the atomic populations determine which is the fraction of the output intensity associated to each polarization component, as given by Eqs.~(\ref{NormIntLargeL}).

To summarize, applying the CRIB technique the phaseonium medium can act as a tunable polarization qubit splitter: part of the field ends up in the antisymmetric (unabsorbed) mode, which exits in the forward direction, while the rest is absorbed. Moreover, the stored antisymmetric mode can be recovered on-demand in the backward direction with, in the ideal case, arbitrary time delay. This system acts as a state filter or qubit preparator, since one could properly adjust the phaseonium state to filter out the non desired polarization components of a particular photon state. The same idea works for the implementation of a quantum sieve, since only a particular desired superposition of the two field components is stored and the rest exits the medium. 

\section{Quantum Memory in a Longitudinal Phaseonium}
\label{sec:QM}

The analysis performed in the previous section is referred to a system with all the atoms prepared in the same coherent superposition of their ground levels. We have seen that the propagation of a pair of pulses through this spatially homogeneous phaseonium leads to an automatic adjusting of the pulse components to fulfill condition (\ref{EITCond}). This prevents complete absorption and, in turn, the straightforward implementation of a quantum memory. In this Section we will see that the latter can nevertheless be overcome by imposing that the preparation of the $\Lambda$-system depends on the position. In particular, we consider that the phase of the two-photon coherence $\phi_{12}$ varies with the position along the light propagation direction:
\begin{eqnarray} \label{PhaseZ}
\phi_{12}(z)=\theta\frac{z}{L},
\end{eqnarray}
with $\theta$ being the phase imposed in the medium at $z=L$. 
This preparation could be implemented, for instance, by applying a linear magnetic field gradient coupling the two ground levels such that it produces an opposite Zeeman shift of their energies that depends on the magnetic field strength. After a certain time the magnetic field is switched off yielding a spatial linear phase between the two ground levels. 
If so, the equations describing the evolution of the system are the same as (\ref{sigma13eq})-(\ref{fieldeqf}), but now with the corresponding position dependence of the two-photon coherence phase. Following the same procedure as in the previous section, an analytical solution for the field components propagating in a position dependent phaseonium medium can be obtained:
\begin{widetext}
% Solucions Absorci Lambda Coherent phi(z)
\begin{eqnarray} \label{SolPhaseZ}
\widetilde{\Omega}^{\rm in}_{\mu\rho}(z,\omega)=e^{(-1)^{\nu} \frac{i \theta z}{2 L}-\frac{\alpha(\omega) z}{2}} \left[\widetilde{\Omega}^{\rm in}_{\mu\rho}(0,\omega)\left(\cosh{\left(\frac{K_{+}z}{2 L}\right)}-(-1)^{\nu} \frac{i \theta}{K_{+}} \sinh{\left(\frac{K_{+}z}{2 L}\right)}\right)-\widetilde{\Omega}^{\rm in}_{\nu\rho}(0,\omega) \frac{\alpha(\omega) L}{K_{+}} \sinh{\left(\frac{K_{+}z}{2 L}\right)}\right], \notag \\
\end{eqnarray}
\end{widetext}
where $K_{+}\equiv\sqrt{(\alpha(\omega) L)^2-\theta^2}$, and for simplicity we have taken the particular case $\sigma_{11}=\sigma_{22}=0.5$. 
Analyzing this expression carefully, one realizes that, in general, it decays to zero for large enough optical depths $\alpha(\omega) L$, and only for $\theta=0$ the field components will not be absorbed. The physical reason for this result is that the field components are not able to adjust their amplitudes and phases to fulfill condition (\ref{EITCond}) as they propagate, since the phase of the optical coherence between the two lower levels is different at each position of the medium. Therefore, the dark-state is not populated and consequently each component of the field is absorbed. In particular, if the variation of the phase along the medium is large compared with the optical depth, i.e., $\theta\gg \alpha(\omega) L$, an exponential decay of each component is observed:
\begin{eqnarray} \label{SolPhaseZlargeC}
\widetilde{\Omega}^{\rm in}_{\mu\rho}(z,\omega)\simeq e^{-\frac{\alpha(\omega) z}{2}} \widetilde{\Omega}^{\rm in}_{\mu\rho}(0,\omega) .
\end{eqnarray}

As in the standard CRIB approach, the stored information can be retrieved in the backward direction by reversing the sign of the detuning and applying a phase matching operation. This leads to a backward propagating solution of the field components of the form:
%
% Solucions Back Reem Lambda Coherent phi(z)
\begin{eqnarray} \label{SolBackPhaseZ}
\widetilde{\Omega}^{\rm b}_{\mu\rho}(z,\omega)=C_{\mu\mu}\widetilde{\Omega}^{\rm in}_{\mu\rho}(0,-\omega)+ C_{\mu\nu}\widetilde{\Omega}^{\rm in}_{\nu\rho}(0,-\omega) ,
\end{eqnarray}
where
\begin{widetext}
\begin{subequations}
\begin{eqnarray} \label{SolBackCoeffs}
C_{\mu\mu}&\equiv&\frac{e^{(-1)^{\nu}i\frac{\theta z}{2L}}\eta J(\omega)}{QK_{-}\eta(F(\omega)+H(-\omega))}%
\left\{e^{\frac{z\eta F(\omega)}{2}}e^{-\frac{L\eta(F(\omega)+H(-\omega))}{2}}\left[%
K_{-}\left[ (-1)^{\nu}i\theta \sinh{(W)}+%
Q \cosh{(W)} \right]\cosh{\left(\frac{K_{-}}{2}\right)}\right.\right. \notag \\%
&+&\left.
\left[(\theta^2+L^2\eta^2F(\omega)H(-\omega))\sinh{(W)}-%
(-1)^{\nu}i\theta Q\cosh{(W)}\right]\sinh{\left(\frac{K_{-}}{2}\right)}\right] \notag \\%
&-&\left.
e^{-\frac{z\eta H(-\omega)}{2}}Q\left[K_{-}\cosh{\left(\frac{K_{-}z}{2L}\right)}-%
(-1)^{\nu}i\theta\sinh{\left(\frac{K_{-}z}{2L}\right)}\right]
\right\} , \\
C_{\mu\nu}&\equiv&\frac{-e^{(-1)^{\nu}i\frac{\theta z}{2L}}\eta J(\omega) L}{QK_{-}\eta(F(\omega)+H(-\omega))}%
\left\{e^{\frac{z\eta F(\omega)}{2}}e^{-\frac{L\eta(F(\omega)+H(-\omega))}{2}}\left[%
K_{-}\eta F(\omega) \sinh{(W)}\cosh{\left(\frac{K_{-}}{2}\right)}\right.\right.\notag \\%
&+&\left.
\left[Q\eta H(-\omega) \cosh{(W)}+(-1)^{\nu}i\theta\eta(F(\omega)+H(-\omega))\sinh{(W)}\right]\sinh{\left(\frac{K_{-}}{2}\right)}\right] \notag \\%
&-&\left.e^{-\frac{z\eta H(-\omega)}{2}}Q\eta H(-\omega) \sinh{\left(\frac{K_{-}z}{2L}\right)}\right\} ,
\end{eqnarray}
\end{subequations}
\end{widetext}
and $K_{-}\equiv\sqrt{(\eta H(-\omega) L)^2-\theta^2}$, $W\equiv Q(1-z/L)/2$ with $Q\equiv\sqrt{(\eta F(\omega) L)^2-\theta^2}$.
These cumbersome expressions can be written in a much simpler form using the assumption $F(\omega)\simeq H(-\omega)\simeq J(\omega)/2\simeq\pi/2$, which leads to $K_{-}\simeq K_{+}\simeq Q$.
Moreover, one can realize that for the case of optically thick medium ($\alpha L\rightarrow\infty$), i.e. when the incident field is completely absorbed, solution (\ref{SolBackPhaseZ}) simplifies to
%
% Solucions Back Reem Lambda Coherent phi(z) Lim c>>LhH
\begin{eqnarray} \label{SolBackPhaseZlargeC}
\widetilde{\Omega}^{\rm b}_{\mu\rho}(z,\omega)=-\widetilde{\Omega}^{\rm in}_{\mu\rho}(z,\omega) .
\end{eqnarray}
Thus, each component of the field is perfectly retrieved with a global phase change of $\pi$. 

%%%%%%%%%%%%%%%%%%%%%%%
\begin{figure} {\includegraphics[width=\columnwidth]{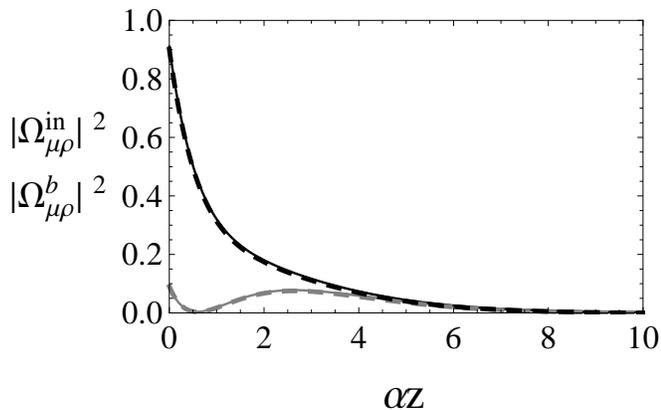}}
\caption{
Normalized intensities of the single pulse polarization components (see text for definition) $\left|\Omega_{13}\right|^{2}$ (black line) and $\left|\Omega_{23}\right|^{2}$ (gray line) as a function of the optical distance $\alpha z$. The solid lines correspond to the incident field while the dashed lines correspond to the backward retrieved components. Initially $\Lambda$-type three-level atoms of phaseonium are prepared in state $\sigma_{11}=\sigma_{22}=0.5$ with phase gradient of $\theta=3\pi$.
The field components are initially weighed as $I^{\rm in}_{13}(0,t_{c})=0.9$, $I^{\rm in}_{23}(0,t_{c})=0.1$ and the relative phase between them is $\phi_{13}(0)-\phi_{23}(0)=0$ both for the forward imput and the backward output.}
\label{f:QMLambdaphiz}
\end{figure}
%%%%%%%%%%%%%%%%%%%%%%%%

Fig.~\ref{f:QMLambdaphiz} shows an example of the absorbed and retrieved normalized intensity of each polarization component $\left|\Omega_{13}\right|^{2}$ (black line) and $\left|\Omega_{23}\right|^{2}$ (gray line) as a function of the optical distance $\alpha z$. This plot clearly shows that the incident polarization components (solid lines), given by Eqs.~(\ref{SolPhaseZ}), are completely absorbed at larger optical distance $\alpha z$. The parameter values considered in Fig.~\ref{f:QMLambdaphiz} are $\sigma_{11}=\sigma_{22}=0.5$, $\theta=3\pi$, $\phi_{13}(0)-\phi_{23}(0)=0$, $I^{\rm in}_{13}(0,t_{c})=0.9$ and $I^{\rm in}_{23}(0,t_{c})=0.1$. The dashed lines correspond to the components of the backward retrieved components, Eq.~(\ref{SolBackPhaseZ}), and we observe practically no difference with the incident components. At $z=0$, the relative phase between the two polarization components, not shown in the figure, is the same for the stored (forward input) and retrieved (backward output) photons. Thus, the system is able to store and retrieve on-demand a single polarization qubit
with unit fidelity.

\section{Numerical Analysis}
\label{sec:Numerics}

The results reported in the previous sections are based on the assumption that the atomic populations and the two-photon coherence $\sigma_{12}$ do not evolve in time, since we have considered the weak field approximation. In order to check the validity of the analytical approach, we have performed exact numerical simulations by integrating the full optical Bloch equations (\ref{sigmaijeqs}) and (\ref{fieldeqf}) with a finite difference method, similar to the one described in \cite{Viscoretal} and assuming Gaussian temporal profiles for the pulse components.

\begin{figure} 
{\includegraphics[width=\columnwidth]{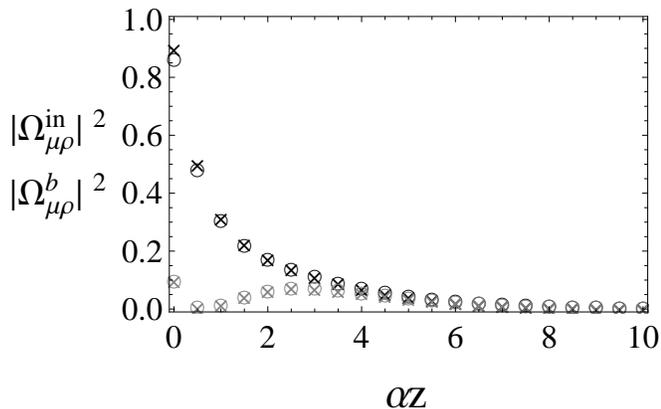}}
\caption{ 
Normalized intensities of the single pulse polarization components coupled with the $\left|1\right\rangle-\left|3\right\rangle$ (black symbols) and $\left|2\right\rangle-\left|3\right\rangle$ (grey symbols) optical transitions as a function of the optical distance $\alpha z$ after numerical integration of the full optical Bloch equations (\ref{sigmaijeqs}) and (\ref{fieldeqf}). The crosses correspond to the propagation of the incident field while the circles correspond to the backward retrieved components. The parameters are the same as in Fig.\ref{f:QMLambdaphiz}.}
\label{f:QMLambdaphizNum}
\end{figure}

An example of the numerical results corresponding to the case of 
the longitudinal phaseonium medium is shown in Fig.~\ref{f:QMLambdaphizNum}.
In this figure, the normalized peak intensity of both field components is shown as a function of the optical length with the same parameter values as in Fig.~\ref{f:QMLambdaphiz}. Comparing these two figures we conclude that the analytical results are in complete agreement with the numerical simulations. Moreover, we have checked that, as expected, the relative phase between the incident field and the backward reemitted field is preserved, and the populations and the two-photon coherence do not exhibit relevant dynamics. This confirms the validity of the weak field approximations performed in the previous sections.

%\FloatBarrier
%\FloatBarrier

\FloatBarrier

\section{Conclusions}
\label{sec:Conclusions}

We have studied the propagation of a single photon pulse whose two
% circular
polarization components are coupled with the two transitions of a coherently prepared $\Lambda$-type three-level medium
presenting artificial inhomogeneous broadening. The propagation effects that normally exhibit this kind of systems have been used in combination with the CRIB technique to discuss potential quantum information applications. On the one hand, we have proposed the use of the $\Lambda$-type system as a quantum filter. This proposal is based on the fact that part of the incident pulse, i.e., the antisymmetric normal mode uniquely determined by the preparation of the atoms in the phaseonium state, propagates without distortion. On the other hand, we have shown that the orthogonal component associated with the symmetric normal mode can be efficiently and completely absorbed and retrieved in the backward direction using the CRIB method. In this case, the system can be used to implement a quantum sieve or, considering both orthogonal modes, a tunable polarization qubit splitter.

Furthermore, we have seen that, by adding a position-dependent phase coherence in the phaseonium medium, the field components can not populate the dark state allowing for a complete absorption of both field components.
Then, by applying the CRIB technique, both components can be recovered on-demand, thus implementing a quantum memory for polarization qubits.

Finally the validity of the analytical approach, which is based on the weak field approximation, has been checked by numerically integrating the full optical Bloch equations. The numerical results obtained are in a very good agreement with the analytical solutions.

\begin{acknowledgments}
A.F. gratefully acknowledges C. Ottaviani for insightful comments and suggestions. We acknowledge support from the Spanish Ministry of Science and Innovation under Contract No. FIS2008-02425, No. HI2008-0238, and No. CSD2006-00019 (Consolider project Quantum Optical Information Technologies), and from the Catalan government under Contract No. SGR2009-00347.
\end{acknowledgments}

\end{document}